# Validation of Fully-Automated Deep Learning-Based Fibroglandular Tissue Segmentation for Efficient and Reliable Quantitation of Background Parenchymal Enhancement in Breast MRI

Yu-Tzu Kuo*[a,b], Anum S. Kazerouni[b], Vivian Y. Park[d], Wesley Surento[b,c], Suleeporn Sujichantararat[b], Daniel S. Hippe[e], Habib Rahbar[b], Savannah C. Partridge[a,b]

[a]Department of Bioengineering, [b]Department of Radiology, and [c]Department of Biomedical Informatics and Medical Education, University of Washington, Seattle, WA, USA; [d]Department of Radiology, Severance Hospital, Research Institute of Radiological Science, Yonsei University College of Medicine, Seoul, KR; [e]Clinical Research Division, Fred Hutchinson Cancer Center, Seattle, WA, USA

## ABSTRACTS

Background parenchymal enhancement (BPE) on breast dynamic contrast-enhanced magnetic resonance imaging (DCE-MRI) shows potential as a breast cancer risk marker. Clinically, BPE is qualitatively assessed by radiologists, but quantitative BPE measures offer potential for more precise risk evaluation. This study evaluated an existing open-source, fully-automated deep learning-based (DL-based) method for segmenting fibroglandular tissue (FGT) to quantify BPE and compared it to a semi-automated fuzzy c-means method. Using breast MRI examinations from 100 women, we evaluated segmentation agreement, concordance across quantitative BPE metrics, and associations with qualitative BPE. The quality of FGT segmentations from both methods was scored by a radiologist. While the DL-based and semi-automated methods showed good agreement for quantitative BPE measurements, DL-based measures more strongly correlated with qualitative BPE assessments and DL-based segmentations were scored as higher quality by the radiologist. Our findings suggest that DL-based FGT segmentation enhances efficiency for objective BPE quantification and may improve standardized breast cancer risk assessment.

## 1. INTRODUCTION

Breast cancer is the second leading cause of cancer-related death among women in the United States, with over 40,000 deaths each year[1]. For women at high risk of developing breast cancer, breast magnetic resonance imaging (MRI) is recommended to supplement annual mammography screenings to facilitate earlier cancer detection and intervention, which can improve treatment outcomes and survival[2,3]. Several imaging features derived from screening examinations have been shown to be associated with breast cancer risk, including mammographic breast density[4,5] and background parenchymal enhancement (BPE) on breast MRI[6]. BPE refers to the normal tissue enhancement observed on dynamic contrast-enhanced MRI (DCE-MRI) and studies have linked high BPE levels to an increased risk of breast cancer[7,8]. Conventional assessment of BPE is performed qualitatively by a radiologist using the Breast Imaging Reporting and Data System (BI-RADS), with four levels: minimal, mild, moderate, or marked[9]. However, visual assessment of BPE is subjective and susceptible to inter-reader variability, which can limit the reliability of this marker for risk stratification. In contrast, quantitative approaches may provide more objective and consistent measurement, enabling the detection of complex imaging patterns and subtle tissue changes while potentially offering a more precise assessment of BPE[10,11].

Accurate extraction of quantitative BPE requires high-quality segmentation of fibroglandular tissue (FGT) to isolate relevant tissue structures and eliminate background or unrelated anatomical components (e.g., pectoral muscles and vessels in the breast region). Fuzzy c-means (FCM) clustering is commonly used for efficient medical image segmentation, as it identifies clusters of voxels with similar signal intensities, making it suitable for distinguishing tissues with varying contrast. However, FGT, vessels, skin, and other structures can exhibit similar signal intensities on breast MRI, leading to segmentation inaccuracies. Additionally, FCM-based segmentation of FGT often involves manual delineation of the breast region and refining the probability threshold for clustering, making it a semi-automated

Corresponding author: Yu-Tzu Kuo (yutzukuo@uw.edu)

segmentation method[12,13]. Although this approach provides whole volume segmentation and facilitates BPE quantitation, its semi-automated nature may introduce inter-operator variability, potentially leading to inconsistencies in FCM outputs. These factors raise concerns regarding the repeatability and reliability of the segmentation process.

Deep learning (DL) based image segmentation offers a promising alternative for improving efficiency and consistency of segmentation, yielding a fully-automated method that eliminates manual interventions and associated interoperator variability[14,15]. U-Net, a widely adopted architecture for image segmentation, integrates the features of a fully convolutional network and can capture complex structures and internal features beyond signal intensities[16]. However, the performance of DL models is highly dependent on the training dataset and may vary when applied to images from different institutions due to scanner differences and variations in acquisition parameters. To evaluate the robustness and generalizability of such models, it is crucial to assess their performance across diverse image datasets.

In this study, we assessed different FGT segmentation techniques to better understand their performance and limitations, with the ultimate goal of creating a more accurate and efficient imaging-based risk assessment for breast cancer. Specifically, we evaluated the robustness and generalizability of an existing open-source DL model for FGT segmentation[15] by applying it to our own institutional data. We compared its performance to that of conventional FCM-based FGT segmentation based on reproducibility and quality of segmentations, as well as agreement in derived quantitative BPE metrics.

## 2. METHODS

*Dataset*

For this analysis, we randomly selected 100 breast MRI examinations from a larger breast MRI screening cohort dataset compiled for breast cancer risk modeling research. The research dataset included breast MRIs from all women who underwent high-risk breast cancer screening at University of Washington between 2005 and 2015, with no cancer diagnosis at least one year following the MRI, and excluding those with prior breast surgery, history of breast cancer, neoadjuvant therapy prior to the MRI, or less than one year of follow-up. Radiologist-assessed qualitative BPE and breast density were extracted from patients' electronic health records. The study was HIPAA compliant and IRB-approved (Hutch IRB #7339), with need for informed consent waived for retrospective clinical imaging and medical record review.

*MRI scan and image preprocessing*

Breast MRI exams were acquired using either a 1.5-T scanner or a 3-T scanner. While scan protocols varied over time, all exams met the American College of Radiology breast MRI accreditation criteria and included a DCE-MRI sequence with multiple T1-weighted fat-suppressed images acquired before and after the injection of a gadolinium-based contrast agent, with the first post-contrast image centered 90 to 120 seconds after injection. Prior to analysis, breast DCE-MRI was first corrected for patient motion across time points with a 2D affine registration and then resampled to an isotropic resolution of 1 mm × 1 mm × 1 mm using custom software (MATLAB R2020a, Natick, MA). The resampled images had a voxel size ranging from 300 × 300 × 124 to 400 × 400 × 218.

*Semi-automated FGT segmentation*

Using pre-contrast DCE-MRI, breast segmentation was performed using signal intensity thresholding and manual removal of the chest cavity to create a mask of bilateral breast regions (Figure 1A). FCM clustering was then applied to masked breast regions of the pre-contrast DCE-MRI, producing two clusters: fat (dark) and FGT with blood vessels (bright) regions. The FCM probability threshold was adjusted on a case-by-case basis to refine the FGT segmentation (Figure 1B). To assess inter-operator agreement, two readers independently performed FCM-based (semi-automated) segmentation (Semi-auto 1 & Semi-auto 2) for all cases.

*Deep Learning based FGT segmentation*

For DL-based FGT segmentation, we employed an open-source DL algorithm for breast FGT segmentation, developed by Lew et al.[15]. Image intensities from pre-contrast DCE-MRI were first capped at the 99.9th and 0.1th percentiles to limit extreme values, then normalized using z-score normalization before being input into two sequential U-Net models for whole breast and FGT/vessel segmentation. The entire 3D image volume was input into the Breast U-Net for bilateral whole breast segmentation (Figure 2A). Subsequently, the original image was paired with the generated whole breast segmentation, divided into 3D sub-volumes of size 96 × 96 × 96, and fed into the FGT-Vessel U-Net to segment FGT and blood vessels separately. The final segmentations were reconstructed using the smallest number of sub-volumes necessary to cover and restore the original image volume size. For voxels appearing in multiple sub-volumes, the final prediction was determined by averaging the predictions from all overlapping sub-volumes (Figure 2B).

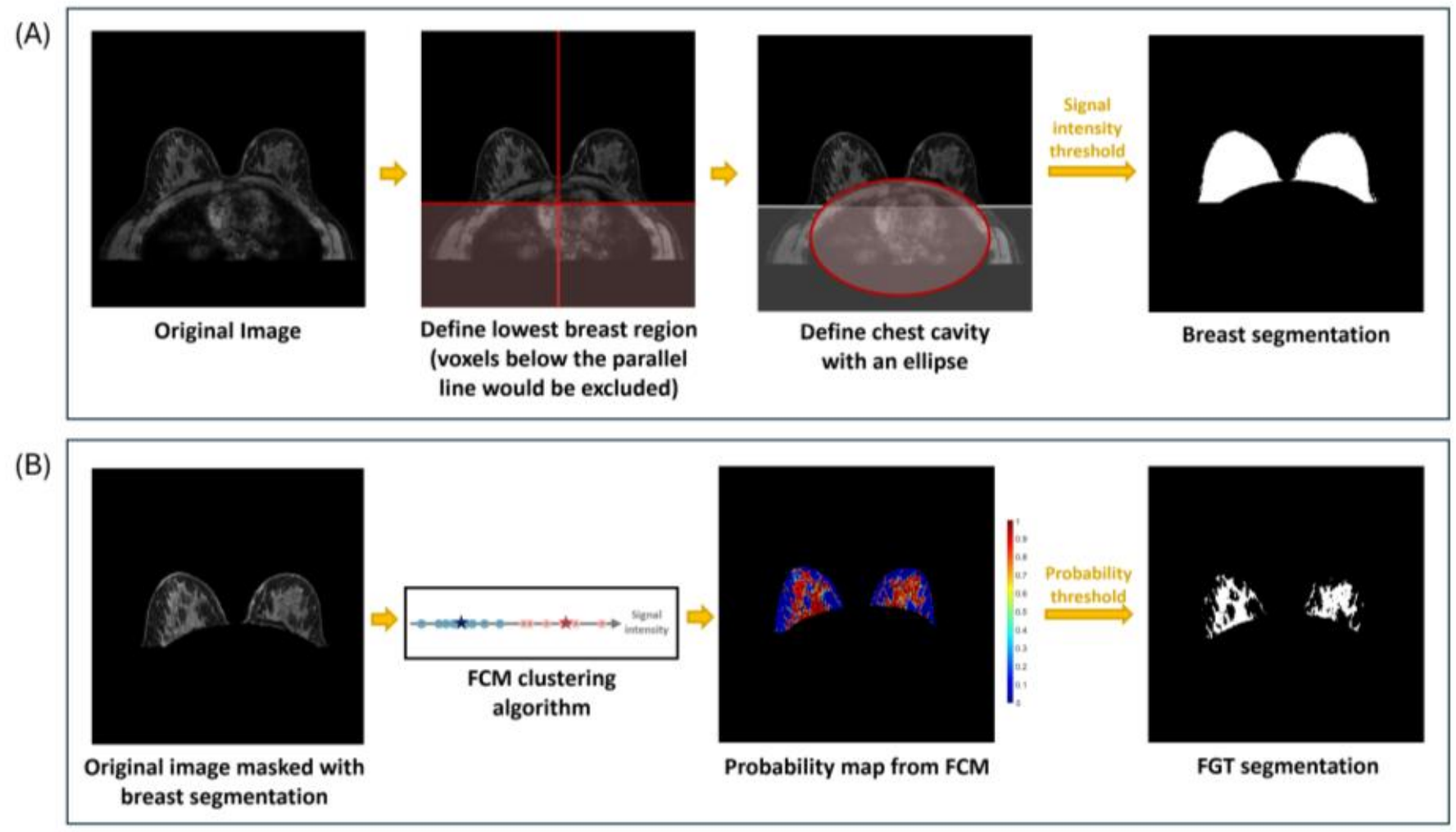

Figure 1. Semi-automated method segmentation process: (A)The breast region and chest cavity were first defined by operators. Whole breast segmentation was then performed using signal intensity thresholding. (B) The image was masked with the breast segmentation and processed using FCM to calculate the probability of each voxel within the segmented breast region, using two clusters representing adipose tissue and FGT. The FCM probability map represented the likelihood of voxels belonging to the higher signal intensity cluster (i.e. FGT). Voxels with a probability exceeding the set threshold were selected to generate the final FGT segmentation.

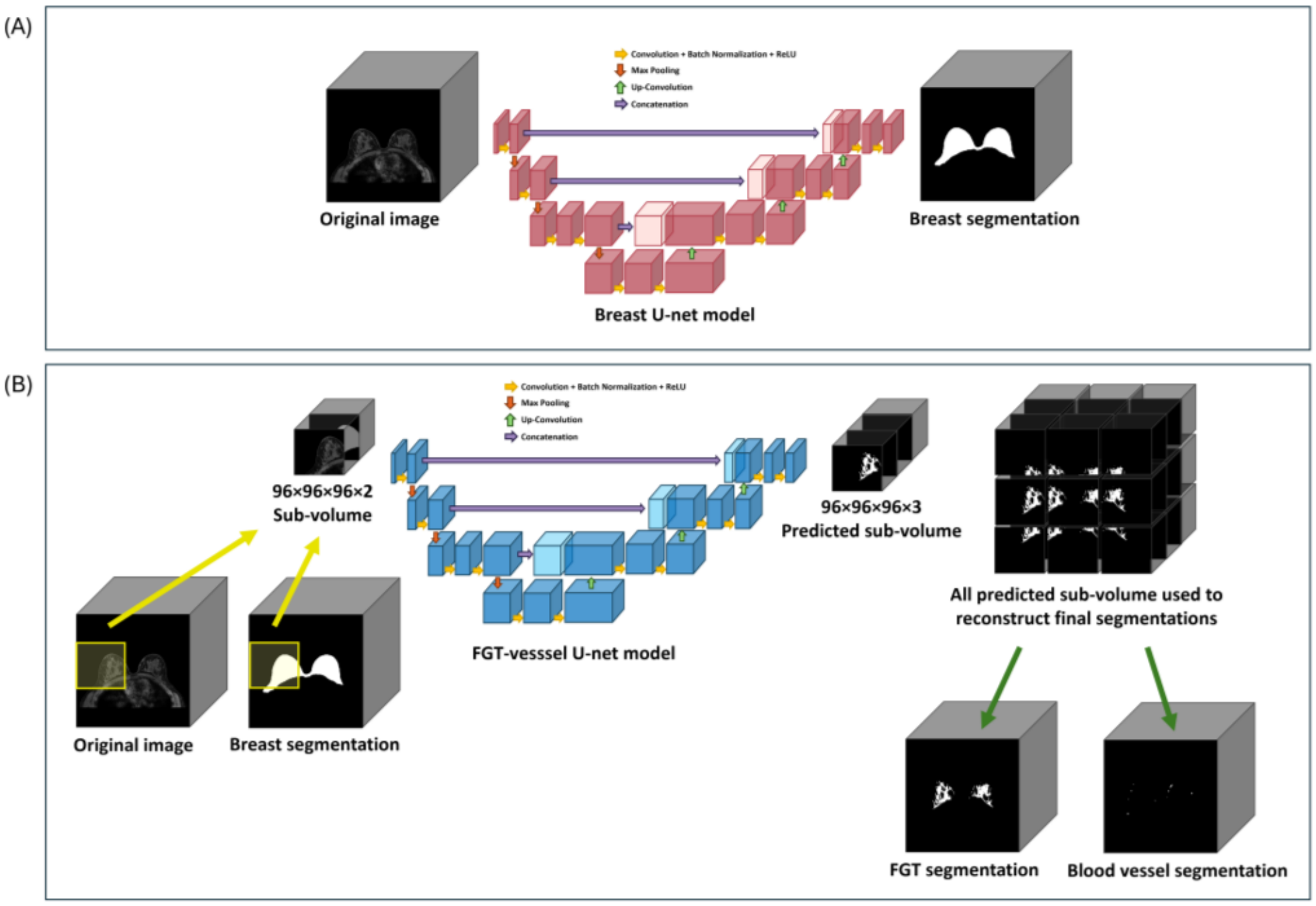

Figure 2. DL method segmentation process: (A) The original image was fed into the Breast U-Net to generate bilateral breast segmentation. (B) The original image and the breast segmentation were then divided into sub-volumes and input into the FGT-Vessel U-Net. Multiple sub-volumes predicted from the same image were processed to reconstruct both the FGT and vessel segmentations.

*Radiologist evaluation of FGT segmentations*

To assess FGT segmentation quality, a reader study was performed where a radiologist with 10 years of subspecialty experience in breast imaging reviewed semi-automated (Semi-auto 1) and DL-based segmentations. The radiologist was presented with the original image volume, alongside FGT segmentations from both methods, with FGT contours displayed in red overlaid on the original breast MRI (Figure 3A). The radiologist was blinded to the segmentation approach for each image pair, with images randomly assigned to the middle or right side of the screen, while the original image was displayed on the left. The quality of each segmentation was scored on a five-point scale based on a visual evaluation of the delineation accuracy of the FGT region across the whole image volume: 1 = unacceptable, 2 = borderline, 3 = clinically acceptable, 4 = good, and 5 = excellent. If any slices were deemed unacceptable, the overall score was restricted to 2 or below. Additionally, If the two segmentation scores for a patient were identical, the radiologist would select the preferred segmentation or indicate no preference (Figure 3B).

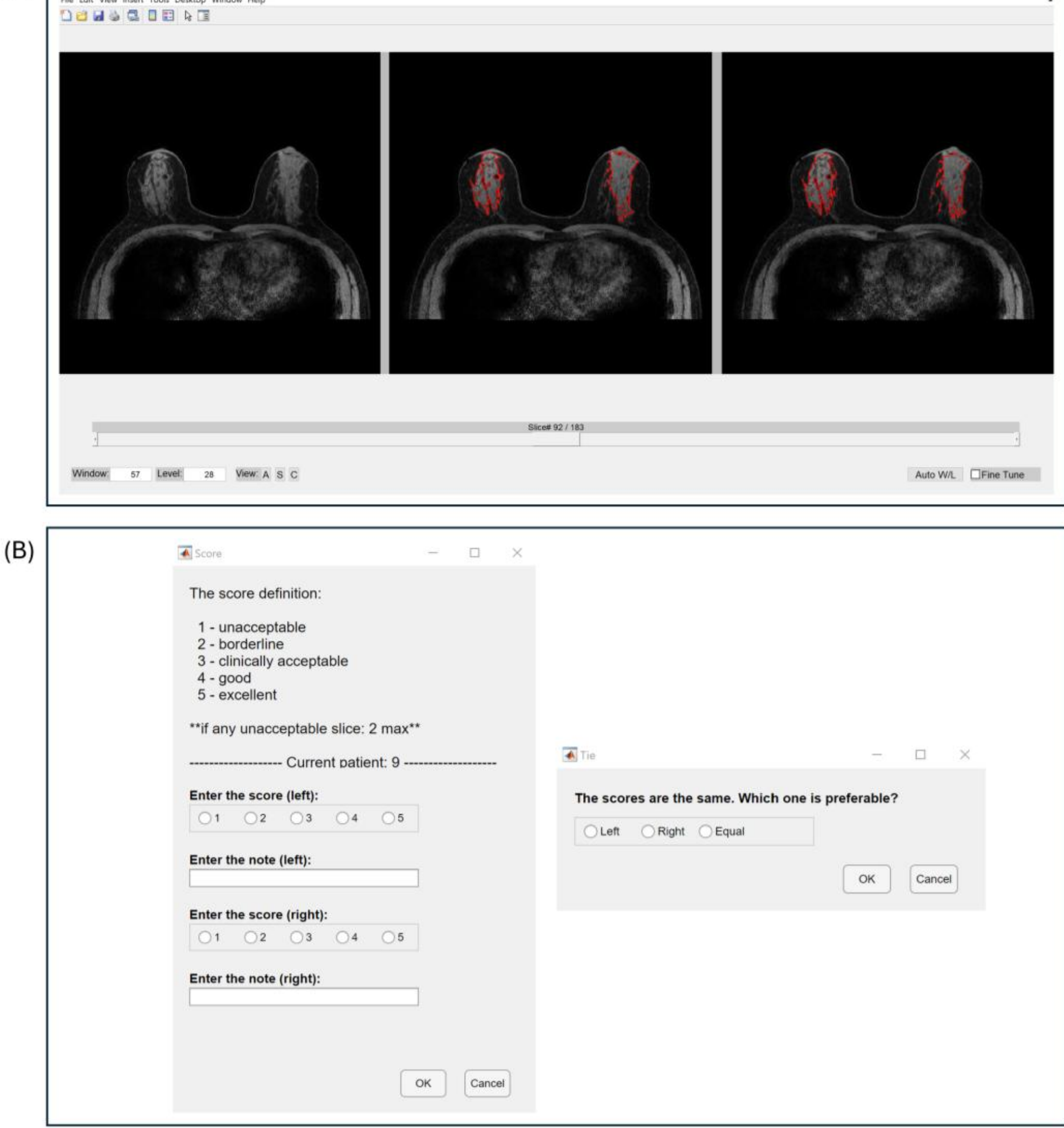

Figure 3. Example reader study set up for the radiologist to evaluate FGT segmentation quality. (A) The original image was displayed on the left, while the two segmentations (DL and Semi-auto 1) were displayed in random order on the middle and right images (red contours depict segmented regions). The radiologist could scroll through slices to view the segmentation results for the entire 3D image volume. (B) The radiologist could then input their segmentation quality scores for the left and right segmentations, and if the scores were equal, the radiologist would further select their preferred FGT segmentation.

*Quantitative BPE analysis*

Percent enhancement (PE) maps were calculated from DCE-MRI using the following formula:

$$PE = \frac{(S1 - S0)}{S0} \times 100\%$$

where S0 and S1 represent the pre- and two minutes post-contrast images, respectively. From these PE maps, four quantitative BPE metrics were measured to characterize extent of BPE and designed to mimic how radiologists qualitatively assess BPE (based on the amount and intensity of normal tissue enhancement). Based on prior work[10,11], the BPE volume was defined as the volume of FGT with a PE value ≥50%. In addition to total BPE volume, three additional metrics were calculated, including the BPE volume: FGT volume ratio, the BPE volume: breast volume ratio, and the BPE integrated intensity, calculated as the BPE volume multiplied by the mean PE of the BPE volume (Figure 4). The concordance correlation coefficient (CCC) was used to assess the consistency of quantitative BPE measurements across segmentation approaches.

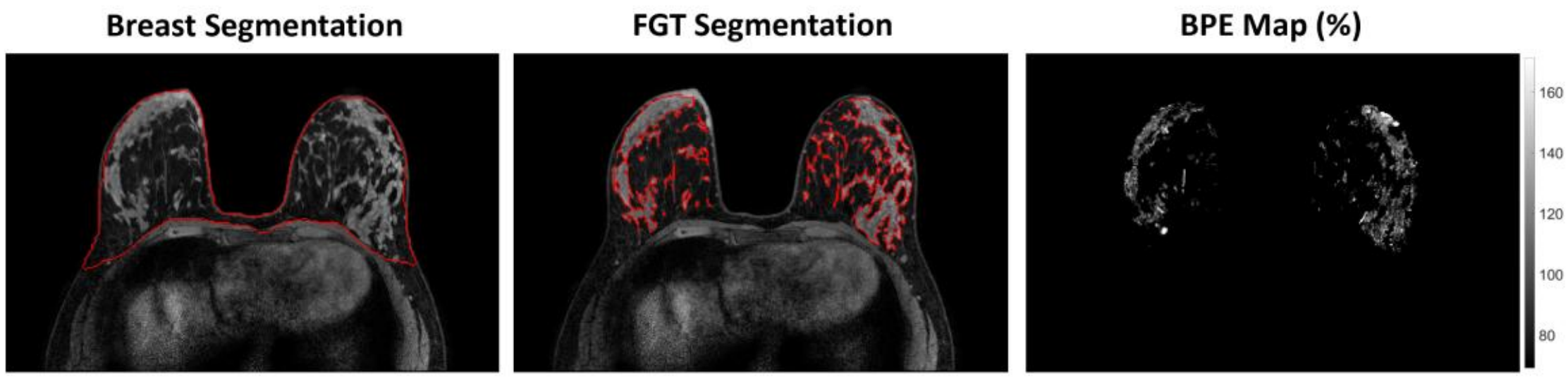

Figure 4. Breast segmentation (left) was used to calculate breast volume. FGT segmentation (middle) was used to calculate FGT volume. BPE map, generated using the FGT segmentation and applying a 50% PE threshold (i.e. only FGT voxels meeting at least 50% enhancement criteria were included), was used to calculate BPE volume.

*Statistical Analysis*

The Dice similarity coefficient (DSC) was used to assess spatial agreement between segmentations to evaluate inter-operator reproducibility for semi-automated segmentations as well as agreement between semi-automated and DL-based segmentations. A DSC value of 0 indicates no overlap between segmentations and 1 indicates complete overlap. We further compared the calculated FGT volume across segmentations using the CCC.

Radiologist segmentation quality scores were compared between DL-based and semi-automated FGT segmentations using the Wilcoxon signed-rank test. Additionally, the correlation between quantitative BPE metrics and qualitative BPE (categorized as 1–minimal, 2–mild, 3–moderate, and 4–marked) was assessed using Spearman's rank correlation coefficient. Spearman's rank correlation coefficients were compared between segmentation techniques using the nonparametric bootstrap.

## 3. RESULTS

*Dataset*

The entire cohort consisted of 100 patients (median age 45.5, range 25 to 77), including 2 with fatty breasts, 25 with scattered density, 50 with heterogeneously dense breasts, and 23 with extremely dense breasts. Qualitative BPE of the cohort included 31 minimal, 30 mild, 21 moderate, and 18 marked BPE cases.

*Agreement and reproducibility between segmentation methods*

Semi-automated FGT segmentations showed excellent reproducibility. Spatial agreement was high between segmentations by the two operators (Semi-auto 1 and 2), with a median Dice coefficient of 0.94 (IQR: 0.89-0.97). The FGT volumes derived from the two semi-automated segmentations demonstrated strong concordance, with a CCC of 0.95 (95% CI: 0.93-0.97). Agreement between the DL-based FGT segmentation and the two semi-automated segmentations was moderate, with a median Dice coefficient of 0.82 (IQR: 0.74-0.88) for Semi-auto 1 and 0.82 (IQR: 0.72-0.88) for Semi-auto 2. The CCC for FGT volume between the DL-based and semi-automated segmentations were 0.60 (95% CI: 0.46-0.71) and 0.63 (95% CI: 0.50-0.73), respectively.

*Radiologist evaluation of FGT segmentation quality*

The radiologist rated the DL-based FGT segmentation significantly higher in quality compared to semi-automated segmentation with mean scores of 4.2 (IQR: 4.0-5.0) and 3.0 (IQR: 2.0-4.0), respectively (P < 0.0001, Figure 5). DL-based segmentation yielded higher quality scores across all breast densities (P < 0.0001, Figure 5), except fatty breasts. Moreover, DL-based segmentations were preferred over semi-automated segmentations in 93% (93/100) of cases, either receiving higher scores or the same scores but being marked as preferable by the radiologist.

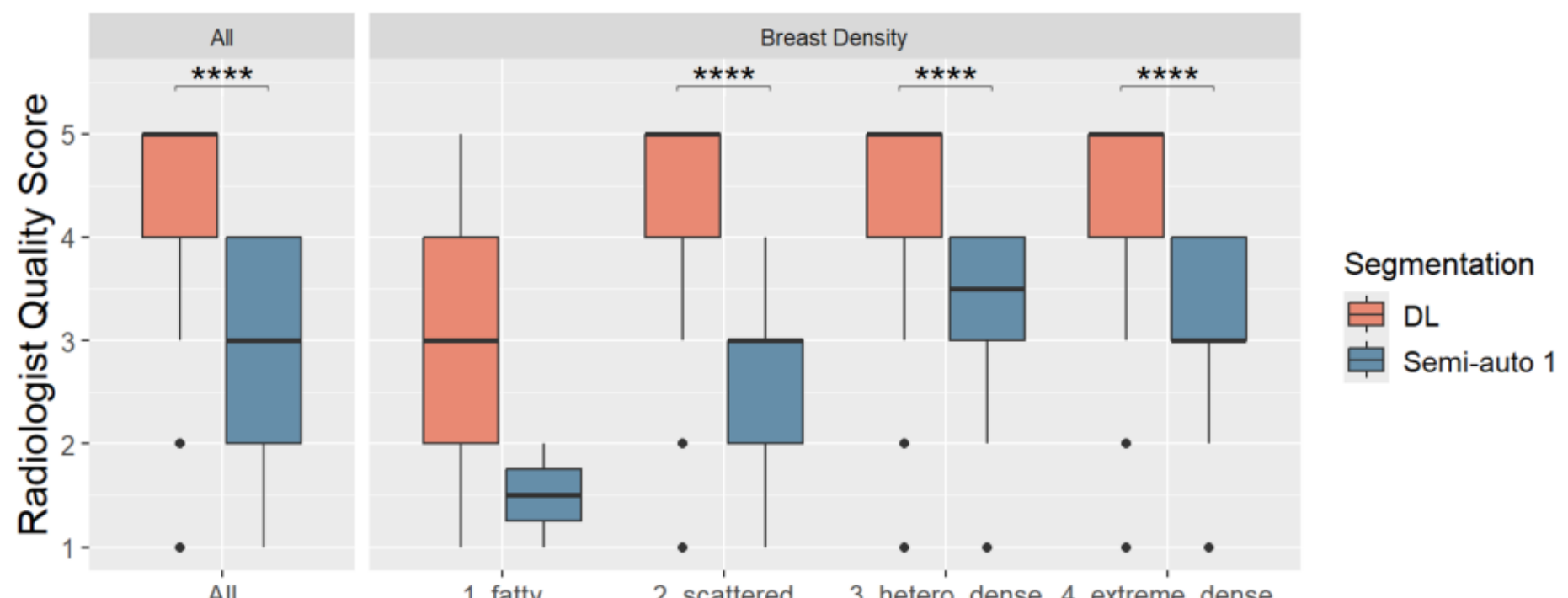

Figure 5. Boxplots showing radiologist-assessed quality scores for DL and semi-automated (Semi-auto 1) FGT segmentations across our cohort ("All") and broken down by breast density category. The Wilcoxon signed-rank test was performed to compare the scores between DL and semi-automated segmentations (**** indicates p-value<0.0001).

Figures 6 and 7 show samples from both segmentation methods, highlighting areas where tissue misclassification may occur. The semi-automated method misclassified blood vessels and regions with inadequate fat suppression that showed similar intensity to FGT (Figure 6B, 6C). Both approaches encountered challenges when segmenting FGT in extremely dense breasts (Figure 6D). Issues in breast segmentation in the semi-automated approach led to occasional inclusion of skin regions (Figure 7A). Similarly, issues in breast segmentation for the DL-based method led to inclusion of muscle regions in the FGT segmentation (Figure 7B). Both segmentation approaches sometimes encountered challenges when encountering images with implants (Figure 7C).

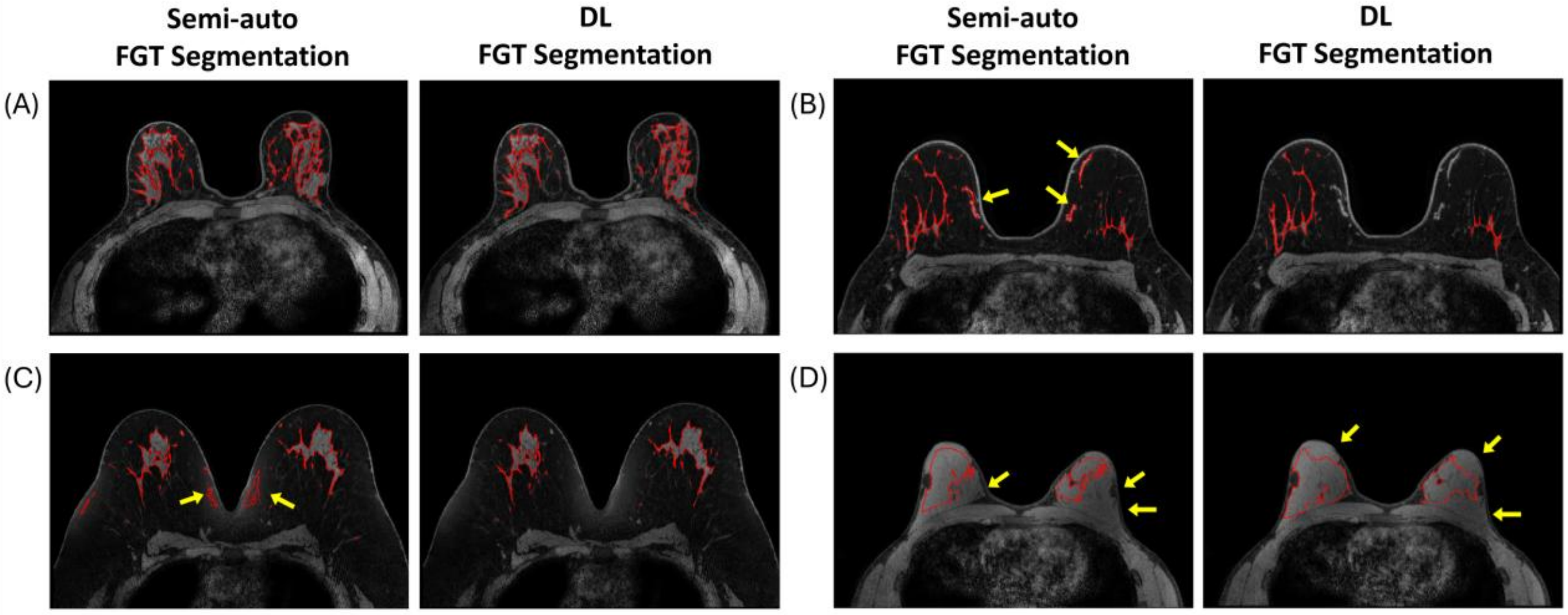

Figure 6. Deep learning and semi-automated segmentation results vary across different cases due to the characteristics of the algorithm: (A) Both methods performed well; (B) The semi-automated method included vessels in FGT segmentation; (C) The semi-automated method included artifactual regions of unsuppressed fat; (D) Both methods had difficulty segmenting extremely dense breast tissue.

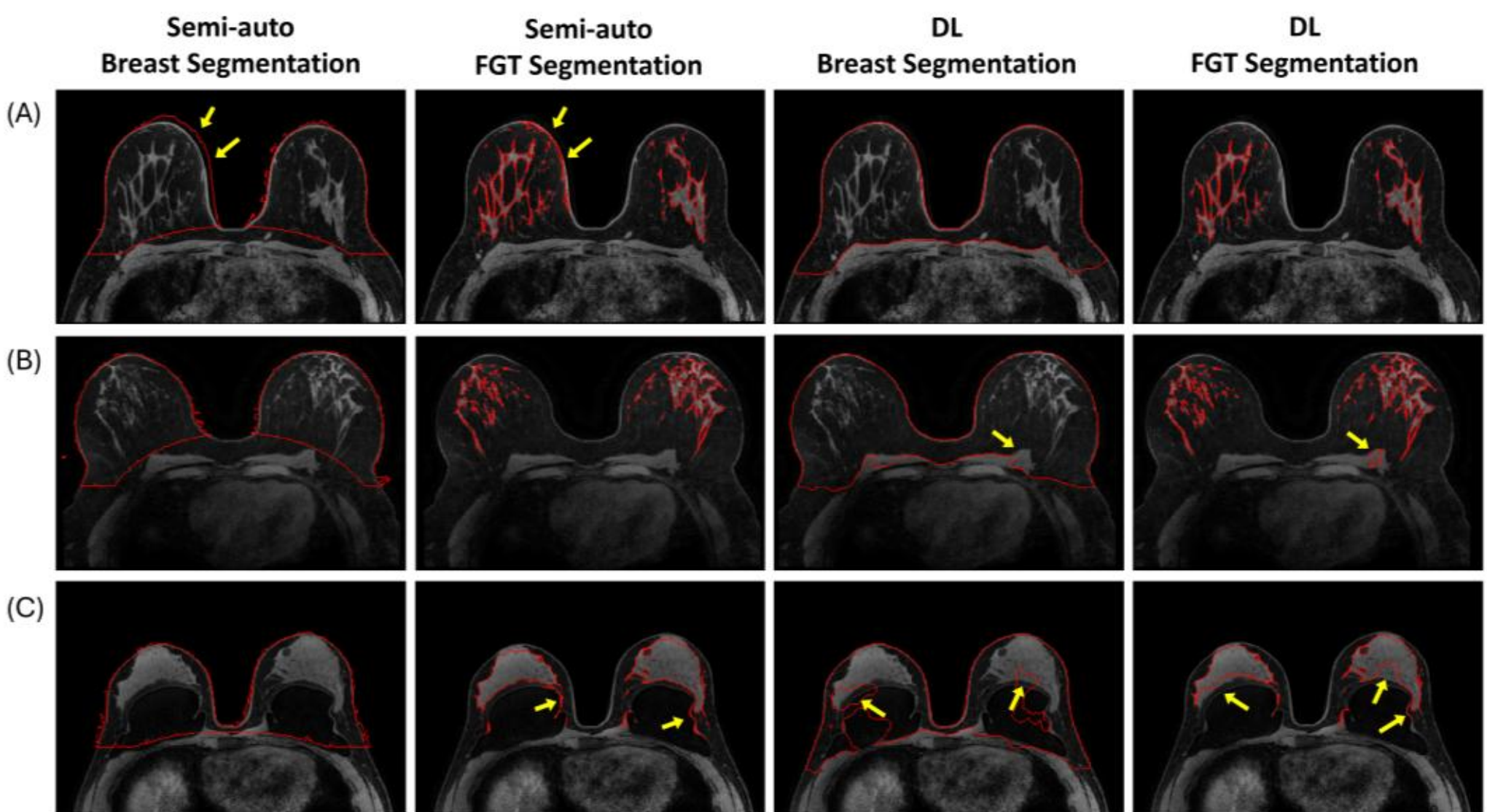

Figure 7. Example cases with breast segmentation issues leading to differences in deep learning and semi-automated segmentation results. (A) The semi-automated method may include skin regions into the FGT segmentation; (B) The DL method may erroneously include pectoral muscle into breast and FGT segmentation; (C) Both methods showed limitations with accurately segmenting breast and FGT in the presence of implants.

*Quantitative BPE analysis*

Table 1 displays the CCC values for quantitative BPE metrics across segmentation approaches. Quantitative BPE metrics exhibited excellent agreement between semi-automated segmentations, with CCC values ranging from 0.98 to 0.99. Similarly, quantitative BPE metrics showed strong agreement between the DL-based and semi-automated segmentation methods, with CCC values ranging from 0.96 to 0.99.

Significant correlations were found between the qualitative and quantitative BPE metrics across all segmentation methods ($P < 0.001$), except for BPE volume: breast volume ratio (Table 2). Notably, no significant differences in correlation coefficients for qualitative BPE and four quantitative BPE metrics were observed between the two semi-automated segmentations ($P > 0.56$ for both). The DL-based segmentation demonstrated significantly higher correlation coefficients for BPE volume, BPE volume: FGT ratio, and BPE integrated intensity compared to Semi Auto 1 ($P < 0.05$) and exhibited a trend toward higher correlations than Semi Auto 2 for the same metrics ($P < 0.1$). However, DL-based segmentation showed a significantly lower correlation coefficient for the BPE volume: breast ratio compared to both semi-automated segmentations ($P < 0.05$).

Table 1. Agreement between quantitative BPE measurements across segmentations.

| **Quantitative BPE metric** | **Concordance Correlation Coefficient (CCC)** | | |
|---|---|---|---|
| | **Semi-auto 1 vs. Semi-auto 2** | **DL vs. Semi-auto 1** | **DL vs. Semi-auto 2** |
| BPE volume ($mm^3$) | 0.98 (0.97-0.99) | 0.96 (0.94-0.97) | 0.96 (0.94-0.97) |
| BPE volume : FGT volume ratio (%) | 0.99 (0.98-0.99) | 0.98 (0.97-0.99) | 0.98 (0.97-0.99) |
| BPE volume : breast volume ratio (%) | 0.99 (0.98-0.99) | 0.98 (0.97-0.98) | 0.97 (0.95-0.98) |
| BPE integrated intensity ($mm^3$) | 0.98 (0.98-0.99) | 0.97 (0.95-0.98) | 0.97 (0.96-0.98) |

*Values in parentheses indicate 95% confidence intervals (CI).

Table 2. Correlation between qualitative and quantitative BPE measurements.

| | Spearman's Correlation | | | | | |
|---|---|---|---|---|---|---|
| | DL | | Semi-auto 1 | | Semi-auto 2 | |
| **Quantitative BPE metric** | **rho** | **P-value** | **rho** | **P-value** | **rho** | **P-value** |
| BPE volume ($mm^3$) | 0.75 | **<0.001** | 0.64 | **<0.001** | 0.65 | **<0.001** |
| BPE volume : FGT volume ratio (%) | 0.62 | **<0.001** | 0.55 | **<0.001** | 0.56 | **<0.001** |
| BPE volume : breast volume ratio (%) | 0.05 | 0.624 | 0.11 | 0.274 | 0.11 | 0.288 |
| BPE integrated intensity ($mm^3$) | 0.74 | **<0.001** | 0.63 | **<0.001** | 0.65 | **<0.001** |

## 4. DISCUSSION AND CONCLUSIONS

Accurate and standardized segmentation of FGT is essential for reliable BPE quantification in breast DCE-MRI, yet existing methods vary in their clinical applicability. This study aimed to evaluate and compare a DL-based automated segmentation approach with a semi-automated method to assess their effectiveness in FGT and BPE quantification. The DL-based segmentation pipeline we implemented (including patch-based reconstruction strategy) is fully-automated, with virtually no operator dependency, and is therefore inherently reproducible across repeated runs. Notably, FGT segmentations generated by the more interactive semi-automated approach were also highly consistent, with minimal inter-operator variability, underscoring the reliability of either approach for assessing BPE. Differences in the segmentation techniques resulted in moderate agreement between the DL-based and semi-automated segmentations. Importantly, radiologist assessment indicated that DL-based segmentations more accurately delineated FGT boundaries and were more often preferred over the semi-automated segmentations, which was observed across different breast density categories. Moreover, the quantitative BPE metrics derived from DL-based FGT segmentations exhibited slightly higher correlation to radiologist qualitative BPE assessments compared to those from semi-automated segmentations.

In cases for which the radiologist assigned a low score for quality, the semi-automated method struggled to differentiate isointense FGT and vessels in pre-contrast DCE-MRI, which is not surprising as FCM relies on intensity-based clustering. Intensity inhomogeneities of unsuppressed fat within the images could also be mistakenly included in the segmentation for the same reason. Voxels capturing skin also exhibited a similar signal intensity to FGT, and if the semi-automated breast segmentation did not exclude these regions initially, they would be included when implementing FGT segmentation with FCM. In contrast, the automated DL-based method was generally able to identify the FGT region without these issues. However, the DL-based method occasionally misclassified the chest wall, muscles, and lymph nodes as FGT due to inaccuracies in DL-based breast segmentation step which included these extraneous tissues. These issues occurred less frequently with the semi-automated method because manual exclusion of the chest cavity and regions outside the breast was performed prior to FGT segmentation. Both segmentation methods faced challenges in cases involving breast implants, where the high-intensity implant boundaries were difficult to distinguish from FGT. Additionally, implants affected the accuracy of breast segmentation in the DL-based method. Neither approach consistently distinguished subtle FGT from surrounding structures, highlighting a common limitation in both methods.

Despite the methodological differences, quantitative BPE metrics showed good agreement between the DL-based and semi-automated methods, suggesting that BPE quantitation is not significantly affected by segmentation discrepancies. The high agreement in quantitative BPE metrics across methods indicates that fully-automated DL-based segmentation can provide reproducible measurements while reducing manual workload. Quantitative BPE metrics, including BPE volume, BPE volume: FGT ratio, and BPE integrated intensity, derived from the DL-based method also suggested a trend of higher correlation with radiologist-assessed qualitative BPE categories than semi-automated segmentation, likely due to more accurate FGT delineation. This implies that deep learning-based segmentation could provide a clinically translatable approach for efficient, objective BPE quantification.

There are some limitations to our study. Our cohort included only a few patients with fatty breast density due to its rarity in our screening population, limiting our ability to compare the performance of both methods on breasts with less FGT. Additionally, since the dataset includes only images acquired before 2015, FGT segmentation performance requires further validation in exams acquired with newer scanner equipment and software, which may affect signal intensity in the images. For the semi-automated segmentation, breast segmentations were relatively coarse, as they were primarily used to exclude inappropriate tissues from FGT segmentations. Also, the chest cavity was excluded using a uniform ellipse-shaped region placed at the same position across the entire 3D MRI volume, which may exclude some breast regions and lead to potential inaccuracies in downstream breast volume calculations in some cases. The moderate correlation between quantitative BPE metrics and qualitative BPE (highest correlation, rho = 0.75) suggests that qualitative BPE assessment is influenced by multiple factors such as distribution, intensity, and pattern of enhancement throughout the breast and possibly patient factors such as breast size and composition. Moreover, no adjustment was made for field strength, which could also influence the quantitative BPE measurements. More sophisticated BPE quantification approaches using radiomics or AI-trained models could be explored to more closely match qualitative assessments.

In summary, while both semi-automated and DL-based segmentation methods produce consistently high-quality segmentation results and demonstrate good reproducibility for quantitative BPE metrics, the DL-based approach appears to provide more accurate FGT segmentation, leading to higher correlation with qualitative BPE assessments. These findings suggest that DL-based segmentation may help standardize BPE measurements, which is critical for integrating BPE into breast cancer risk assessment in clinical care.

## ACKNOWLEDGEMENTS

This research is supported by funding from NIH/NCI U01CA152637 and K99CA293004.